\documentclass{article}

\usepackage{arxiv}

\usepackage[utf8]{inputenc} 
\usepackage[T1]{fontenc}    
\usepackage{lmodern}        
\usepackage{hyperref}       
\usepackage{url}            
\usepackage{booktabs}       
\usepackage{amsfonts}       
\usepackage{nicefrac}       
\usepackage{microtype}      
\usepackage{graphicx}

\title{Estimating Treatment Effects with Missings Not At Random in the
Estimand Framework using Causal Inference.}

\author{
    Aleix Ruiz de Villa
   \\
    Independent Researcher \\
   \\
  \texttt{\href{mailto:aleixrvr@gmail.com}{\nolinkurl{aleixrvr@gmail.com}}} \\
   \And
    Llorenç Badiella
   \\
    Departeament de Matemàtiques \\
    Universitat Autònoma de Barcelona \\
   \\
  \texttt{\href{mailto:llorenc.badiella@uab.cat}{\nolinkurl{llorenc.badiella@uab.cat}}} \\
  }

\providecommand{\tightlist}{%
  \setlength{\itemsep}{0pt}\setlength{\parskip}{0pt}}

\NewDocumentCommand\citeproctext{}{}

\makeatletter
 \let\@cite@ofmt\@firstofone
 \def\@biblabel#1{}
 \def\@cite#1#2{{#1\if@tempswa , #2\fi}}
\makeatother
\newlength{\cslhangindent}
\newlength{\csllabelwidth}
\newenvironment{CSLReferences}[2] 
 {\begin{list}{}{%
  \setlength{\itemindent}{0pt}
  \setlength{\leftmargin}{0pt}
  \setlength{\parsep}{0pt}
  \ifodd #1
   \setlength{\leftmargin}{\cslhangindent}
   \setlength{\itemindent}{-1\cslhangindent}
  \fi
  \setlength{\itemsep}{#2\baselineskip}}}
 {\end{list}}
\usepackage{calc}

\usepackage{amsmath}
\usepackage{graphicx}
\usepackage{float}
\usepackage{centernot}
\usepackage{subcaption}
\usepackage{amsthm}
\usepackage{xcolor}
\begin{document}
\maketitle

\begin{abstract}
The analysis of randomized trials is often complicated by the occurrence
of intercurrent events and missing values. Even though there are
different strategies to address missing values it is still common to
require missing values imputation. In the present article we explore the
estimation of treatment effects in RCTs from a causal inference
perspective under different missing data mechanisms with a particular
emphasis on missings not at random (MNAR). By modelling the missingness
process with directed acylcic graphs and patient-specific potential
response variables, we present a new approach to obtain an unbiased
estimation of treatment effects without needing to impute missing
values. Additionally, we provide a formal that the average conditional
log-odds ratio is a robust measure even under MNAR missing values if
adjusted by sufficient confounders.
\end{abstract}


\newtheorem{assumption}{Assumption}
\newtheorem{theorem}{Theorem}
\newtheorem{lemma}{Lemma}
\newtheorem{corollary}{Corollary}
\newtheorem{definition}{Definition}

\section{Introduction}\label{introduction}

Randomized clinical trials (RCT) remain the gold standard to evaluate
the efficacy and safety of new treatments. Randomization provides the
basis for drawing statistically valid causal inferences for treatment
effect (Qu et al., 2023). However, the analysis of randomized trials is
often complicated by the occurrence of certain intercurrent events
(ICEs) that affect the interpretation of the treatment effect or
preclude the observation of the outcome of interest.

In 2010 with the publication of the US National Research Council report
on `Prevention and Treatment of Missing Data in Clinical Trials',
missing data and sensitivity analyses in RCT came to the forefront. In
2017, the ICH E9 addendum R1 (European Medicines Agency, 2017)
introduced the `estimands framework', an structured approach to be
considered in clinical trials to clearly define the treatment effect a
study aims to estimate. This framework helps to ensure that the research
questions are well-defined and that appropriate study methods are used
to answer these questions, and at the same time aligns the study design,
data collection, and analysis methods with the scientific questions of
interest (Kahan et al., 2024).

To define an estimand, several components must be specified: the
population of interest, the primary endpoint, a measure of intervention
effect, the statistical method to summarise this measure and finally how
intercurrent events will be handled. Intercurrent events often lead to
missing data (patients may discontinue treatment, switch medications, or
miss scheduled visits, resulting in incomplete data collection), in this
sense, these strategies must clearly explain how missing values are
addressed.

Missing data has become an incessant problem in clinical trials that can
bias treatment group comparisons, inflate rates of false negative and
false positive results and compromise the study's integrity.
Fortunately, missing data has been an active area of investigation with
many advances in statistical theory and in the ability to implement that
theory (Mallinckrodt et al., 2017). Rubin (Rubin, 2004) developed a
framework for addressing missing data and described different
missing-data mechanisms. Three types of missing data mechanisms are
commonly described (Van Buuren \& Van Buuren, 2012):

\begin{itemize}
\tightlist
\item
  If the probability of being missing is the same for all cases, then
  the data are said to be \textbf{missing completely at random} (MCAR).
  The causes of the missing data are unrelated to the data. We may
  consequently ignore many of the complexities that arise because data
  are missing, apart from the obvious loss of information. While
  convenient, MCAR is often unrealistic for the data at hand.
\item
  If the probability of being missing is the same only within groups
  defined by the observed data, then the data are \textbf{missing at
  random} (MAR). MAR is a much broader class than MCAR. Modern missing
  data methods generally start from the MAR assumption.
\item
  If neither MCAR nor MAR holds, then we speak of \textbf{missing not at
  random} (MNAR). MNAR means that the probability of being missing
  varies for reasons that are unknown to us. MNAR is the most complex
  case. Strategies to handle MNAR are to find more data about the causes
  for the missingness, or to perform what-if analyses to see how
  sensitive the results are under various scenarios.
\end{itemize}

When some data are missing, it is important that approaches used to deal
with them at the analysis stage align with the strategies chosen to
handle each specific ICE that preceeded missingness. The estimands
framework provides several strategies to handle intercurrent events, and
their associated potentially missing values (Mallinckrodt et al., 2020).

\begin{itemize}
\item
  Treatment Policy Strategy: Under this strategy, the treatment effect
  targeted by the estimand is a combined effect of the initial
  randomized treatment and treatment modified as a result of the ICE,
  thus the ICE is incorporated as part of a treatment policy strategy.
  The occurrence of the ICEs is irrelevant, the data is collected for
  the variable of interest and used regardless of whether or not the ICE
  occurs. It is generally preferred by many regulatory agencies. All
  relevant data for the estimand can still be collected after these ICEs
  and in principle it will not be missing unless the patient withdraws.
\item
  Composite Strategy: The outcome and the occurrence of the intercurrent
  event are combined into a single endpoint. Hence, post-ICE
  measurements are not relevant. However, this method needs
  clarification if the outcome is a quantitative measure.
\item
  While on treatment strategy: The outcome measurements are taken only
  up to the occurrence of a particular ICE. Hence, the ICE impacts the
  duration of treatment, but the relevant data for the estimand will in
  principle be available. Post-ICE data is irrelevant. Therefore, the
  while-on-treatment strategy yields no missing data due to the
  correspondent ICEs if outcome measures are available when the ICE
  occurs.
\item
  Hypothetical Strategy: This approach considers what the outcome would
  have been if the intercurrent event had not occurred. In such
  situation the outcome of interest cannot be observed and may need to
  be implicitly or explicitly predicted or imputed.
\item
  Principal Stratum Strategy: The estimand population is redefined to
  include only patients who would not (or would) experience the
  intercurrent event. The principal stratum strategy modifies the
  population attribute of the estimand.
\end{itemize}

Missing data can be minimized by trial design and conduct; however,
missing data is not completely avoidable regardless of which strategies
are used for ICEs. Even though the aforementioned strategies try to
avoid the occurrence of missing values, in practice, still some data
will be lost for reasons related or unrelated to the ICEs. Most of the
strategies to incorporate ICEs in the estimand will require the
imputation of missing values under different assumptions, and at the
same time, it will be required to assess these assumptions using
sensitivity analyses.

In this paper, we explore the estimation of treatment effects in RCTs
from a causal inference perspective, under different missing data
mechanisms with a particular emphasis on MNAR scenarios. We present a
new approach suited to obtain unbiased estimation of treatment effects
without needing to impute missing values. Specifically, we propose a
patient-specific questionnaire designed to elicit information sufficient
to obtain unbiased estimates even in the presence of unobserved
confounders. Treatment effects will be quantified using the Average
Treatment Effect (ATE) and the average conditional log-odds ratio
(AC-LOR) (see Jun \& Lee (2023), Colnet et al. (2023), Zhang (2009),
Pang et al. (2016) and Karlson \& Jann (2023)). AC-LOR has been shown,
Prentice \& Pyke (1979) and Bartlett et al. (2015), to be a robust
measure of treatment effect in the presence of unobserved confounding in
observational studies when using logistic regression. In this paper, we
provide a non-parametric proof that the AC-LOR remains a robust measure
under MNAR missing data mechanisms when adjusted for sufficient
confounders.

Our main contributions are as follows:

\begin{itemize}
\tightlist
\item
  We establish a connection between the estimand framework and causal
  inference by representing RCTs through graphical models under various
  missing data mechanisms.
\item
  We introduce a patient-informed approach that mitigates the impact of
  unobserved confounding.
\item
  We derive unbiased estimators for both the ATE and the AC-LOR in the
  presence of MNAR missing data.
\item
  We provide a formal proof that the AC-LOR remains a robust measure
  under MNAR mechanisms when adjusted for sufficient confounders.
\end{itemize}

The rest of the paper works as follows. Section 2 connects causal
inference to the estimand framework and missing data processes. Section
3 introduces our approach for MNAR scenarios and discusses practical
approaches to estimation. Section 4 presents simulations, and Section 5
discusses implications and limitations.

\section{Causal inference in the estimand
framework}\label{causal-inference-in-the-estimand-framework}

The addendum R1 states that the study objective (estimand) should be
based on the quantification of how does the outcome of a treatment would
compare to what would have happened to the same subjects under different
treatment conditions. Until recently, relatively little has been
published on the topic of estimation of estimands from the perspective
of modern casual inference. Indeed, perhaps surprisingly, the ICHE9
addendum itself does not explicitly mention causal inference concepts or
methods, although these are clearly relevant (Olarte Parra et al.,
2023). Indeed, ICEs are inherently tied to causality and the
interpretation of results. Depending on clinical perspective, they may
represent part of the treatment, mediators, confounders or outcomes in
their own right (Ratitch et al., 2020). In this sense, a causal
inference approach provides clarity in the design and analysis of
studies and helps clarify causal assumptions.

\subsection{Recent Advances in Causal Graphical Models for Missing
Data}\label{recent-advances-in-causal-graphical-models-for-missing-data}

The last decade has seen rapid progress in the use of causal graphical
models to formalize, diagnose and---when possible---resolve problems of
missing data. One of the essential tools in modern causal inference
approaches is the representation of the relationships between variables
by means of directed acyclic graphs (DAGs) (Tennant et al., 2021). By
illustrating the pathways through which causal effects propagate, DAGs
help in understanding both direct and indirect effects. Additionally,
they aid in deriving different estimators under various sets of
assumptions, ensuring that the chosen methods are robust and appropriate
for the study's goals. In Lee et al. (2023), they review the traditional
MCAR/MAR/MNAR framework for handling missing data and argues that it
often oversimplifies real-world scenarios involving multiple incomplete
variables. The authors propose a more practical, assumption-driven
approach to analysis planning that explicitly links the study context,
missingness mechanisms, and analytical strategies to improve
transparency and validity in epidemiological research.

Mohan et al. (2013) introduces missingness graphs (m-DAGs) as a formal
graphical language for representing the mechanisms that generate
missingness and for encoding dependencies between missingness and
measured variables. The paper formulates the recoverability
(identifiability) problem for a broad class of queries, provides
graphical criteria for when certain quantities can be consistently
estimated from observed (censored) data, and shows how standard
MCAR/MAR/MNAR notions appear naturally in the graphical language. Their
work is also explained in a pedagogical way in Mohan (2022) and extended
in Mohan et al. (2018). Based on their work, Holovchak et al. (2024)
presents a longitudinal case study applying m-DAG recoverability ideas
to a pharmacological trial.

Recent work has substantially expanded the theory and practical scope of
graphical methods for missing data. Nabi et al. (2020) established
completeness results for identifying the full-data law in graphical
models, delineating which MNAR mechanisms are fundamentally
identifiable. Building on this, Bhattacharya et al. (2019) proposed a
general identification algorithm that extends the causal ID framework to
missing-data settings, broadening the class of recoverable distributions
and providing constructive procedures for deriving estimands.
Complementing these advances, Srinivasan et al. (2023) introduced
``entangled missingness'' graphs to represent dependencies in
missingness across clustered or networked units, thereby extending
identifiability concepts beyond independent-data contexts. Finally, Nabi
\& collaborators (2022-\/-2024) developed sensitivity-analysis and
goodness-of-fit tools for non-ignorable missingness, linking graphical
identification theory to practical robustness assessments and applied
model evaluation.

Recent research has extended causal inference methodologies to address
missing data challenges within randomized and hybrid trial designs.
Colnet \& colleagues (2024) provide a comprehensive review of strategies
for integrating randomized and observational data, emphasizing
identification frameworks and estimation techniques that accommodate
incomplete covariate information and enhance external validity. Nugent
et al. (2024) develop causal models and targeted estimators for trials
with partial clustering or imbalanced dependence structures,
illustrating how complex trial designs interact with missingness
mechanisms and affect estimation efficiency. Complementing these
methodological advances, Cornelisz et al. (2020) analyze missing
outcomes in RCTs through a causal-inference framework, arguing for
MNAR-aware analyses and introducing the Random Forest Lee Bounds
approach to obtain bounded treatment-effect estimates under selective
attrition. Together, these studies demonstrate the growing integration
of causal and graphical principles into practical solutions for handling
incomplete data in contemporary clinical and experimental research.
Finally, Heng et al. (2025) examines outcome missingness in the
design-based (randomization-based, finite-population) causal inference
setting. The authors propose a general missingness mechanism, develop an
``imputation and re-imputation'' framework for conducting exact
randomization tests even under misspecified imputation models or
unobserved covariates/interference, and extend the framework to
covariate adjustment and construction of valid finite-population
confidence regions with missing outcomes.

\subsection{A general DAG for RCTs}\label{a-general-dag-for-rcts}

\begin{figure}[ht!]
  \centering
  \includegraphics[scale=0.5]{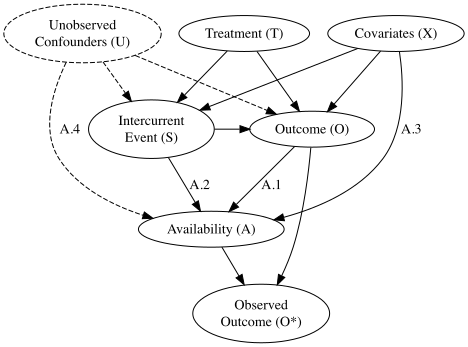}
  \caption{Graph depicting the process of an RCT with a particular incurrent event of interest.}
  \label{fig:dag_general}
\end{figure}

The DAG in figure \ref{fig:dag_general} describes a general process of a
RCT. Each arrow represents a potential causal relationship between two
variables. In particular, an arrow also represents the situation in
which there is no causal relationship. The lack of an arrow represents a
type of independence between variables, and can be considered as an
assumption. For simplicity, the DAG only presents a single intercurrent
event, other events could also be considered, with a similar
configuration regarding arrows.

In a RCT, the treatment \(T\) is randomized, and consequently it doesn't
have any parents. The treatment \(T\) affects the patient's outcome
\(O\). At the same time, treatment could produce a particular side
effect or other intercurrent event (\(S\)). Note that this arrow does
not exclude the situation in which \(S\) is unrelated to the treatment.
The occurrence of \(S\) will presumably affect the patient's outcome
\(O\).

Observed covariables that could affect the outcome such as age, clinical
condition, etc. are denoted as \(X\). These covariates could also be
related to the occurrence of the intercurrent event, and thus they play
the role of confounders between \(S\) and \(O\). There could be also a
group of unobserved covariables \(U\), for instance economic status or
family support, affecting both \(S\) and \(O\) at the same time, and
playing the role of unobserved potential confounders between \(S\) and
\(O\).

The intercurrent event and the outcome may affect the availability and
observation of the outcome: the variable \(O^*\) denotes the observed
outcome variable with potentially missing values. Variable \(A\)
(availability) denotes whether \(O\) has a missing value or not (\(A=0\)
means the value is missing).

\begin{equation}
  O^* =
  \begin{cases}
    O & \text{if}\ A=1 \\
    ? & \text{if}\ A=0
  \end{cases}
\end{equation}

We use the same decomposition of missing values as done in (Mohan et
al., 2013). That is, given a variable \(O\) without missing values, we
denote by \(O^*\) the corresponding observed variable with potential
missing values, and by \(A\) a binary variable that indicates whether
the value of \(O\) is observed (available) or not. Through the paper we
will do a slight abuse of notation. Whenever \(O\) and \(O^*\) appear in
the same formula, we will assume they take the same value. For example,
the expression \(P(O|A=1) = P(O^*|A=1)\) means that for a particular
value \(o\), \(P(O=o|A=1) = P(O^*=o|A=1)\). Additionally, since \(O^*\)
only takes values when conditioning on \(A=1\), we will obviate the
conditioning on \(A=1\) in the notation, and we will write \(O^*\)
instead of \(O|A=1\), such as in \(P(O|A=1) = P(O^*)\).

The reason to have a missing value depends on the intercurrent event
\(S\), the outcome \(O\) and potentially the particular characteristics
of the patient, described by \(U\) and \(X\). For instance, older
individuals may be more sensitive to side effects and, as a result, are
more likely to withdraw from the trial compared to younger participants.

Bornkamp et al. (2021) stated that if the intercurrent event is a
consequence of treatment (possibly generating missing values),
randomization alone is no longer sufficient to meaningfully estimate the
treatment effect. The DAG in figure \ref{fig:dag_general} allows us to
explain why this statement holds. The presence of unobserved confounders
between the ICE occurrence and the missing indicator implies that the
type of missing mechanism is MNAR and this is a challenging situation to
estimate the treatment effect of interest.

In the next sections, we review concepts from causal inference to
characterize precisely the conditions under which standard estimands can
be estimated from trial data under the presence of different missing
values mechanisms. We also present an original approach that consists in
adapting the study design and offers a way to estimate the treatment
effect in the general situation.

\subsection{Estimation of RCT's treatment effects with missing
data}\label{estimation-of-rcts-treatment-effects-with-missing-data}

In the present section, we discuss the estimability of the treatment
effect depending on different assumptions with respect to the missing
data mechanism. More specifically, we will estimate the quantities
\(P(O|T=1), P(O|T=0)\) and the Average Treatment Effect (ATE),
\(P(O|T=1) - P(O|T=0)\), under different missing data mechanisms. Such
mechanisms are described by the presence or not of the arrows denoted as
A.1, A.2 A.3 or A.4 in figure \ref{fig:dag_general}.

\subsubsection{MCAR}\label{mcar}

Under MCAR missing outcome, missing values are generated completely at
random. Thus, it is assumed that there are no causes related to the
missing indicator \(A\).

\begin{figure}[ht!]
  \centering
  \includegraphics[scale=0.5]{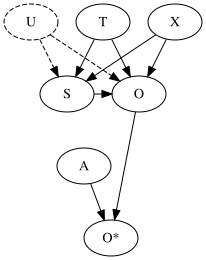}
  \caption{Graph depicting the process of an RCT where missing values are MCAR.}
  \label{fig:mcar}
\end{figure}

This situation is described in figure \ref{fig:mcar}.

The arrows denoted as A.1, A.2, A.3 and A.4 are omitted. In absence of
any row reaching the node \(A\), the missingness mechanism can be
assumed to be MCAR. In this case, the treatment effect can be directly
estimated from the available data, and the estimate of the treatment
effect will be unbiased.

An example of this scenario where missing values in the outcome variable
are produced purely at random would be lost to follow up by
administrative reasons (protocol related, personal circumstances, study
logistics, patients lost to follow up, etc.). It also corresponds to the
situation where the intercurrent does not produce missing values (or
invalid results) and the outcome is available for all patients.

In practice, most of the strategies to deal with intercurrent events
(Treatment Policy, Composite or While on Treatment) are intended to
transform the general DAG (figure \ref{fig:dag_general}) into the
present DAG (figure \ref{fig:mcar}) by adapting the definition of
treatment or the definition of the outcome. These approaches can be used
whenever a valid outcome is available at the moment the intercurrent
event occurs.

Other strategies adapt the definition of the population, trying to
remove the occurrence of the ICE (Principal Stratum Strategy) or
imputing values assuming the ICE did not occur (Hypothetical Strategy).

\subsubsection{MAR}\label{mar}

\begin{figure}[ht!]
  \centering
  \includegraphics[scale=0.5]{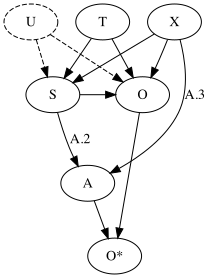}
  \caption{DAG describing a MAR type of missing data. }
  \label{fig:dag_mar}
\end{figure}

When the missing data is MAR, it is assumed that the missing values are
only related to other observed variables. The DAG in figure
\ref{fig:dag_mar} shows this situation, where the only arrows reaching
the node \(A\) are related to observed variables \(S\) and/or \(X\) .

This situation applies when side effects or other intercurrent events
related to safety and produce invalid outcomes or missing values.

In practice, for the analysis of these type of data it is common to
impute the missing values by assuming a particular (usually
conservative) pattern, and to consider different sensitivity analysis to
evaluate the reasonability of the assumptions Mallinckrodt et al.
(2020). There are also other methods in the statistical literature that
can be used to estimate the treatment effect without bias, such as
Maximum Likelihood (Allison, 2012).

From a causal inference point of view, in the MAR scenario, the
treatment effect can be estimated without bias using probabilistic
methods (see RUBIN (1976), Little \& Rubin (2002),Mohan et al. (2013)).
In the following theorem, we present an estimator specifically tailored
to the clinical trial context with intercurrent events.

\begin{theorem}\label{theorem:mar_basic_estimation}
Consider a process described by Figure \ref{fig:dag_mar}. Assume that $X$ and $S$ are categorical variables. Then

$$E(O|T=t) = \sum_{s, x; P(T=t, s, x, A=1) > 0} E(O^*| T, s, x)P(s, x| T=t)$$
\end{theorem}

You can find the proof in \ref{proof:mar_basic_estimation}. Intuitively,
this means that under MAR, conditioning on all observed causes of
missingness suffices to recover unbiased treatment effects using
observed data only. This result relies on the fact that under a MAR
assumption, the availability indicator A is conditionally independent of
the outcome O after conditioning by S and X.

This relation implies that under the MAR assumption, the treatment
effect and the differences between treatments can be estimated without
bias using only the available data as a weighted average of the
treatment effect in different subgroups of subjects and there is no need
to impute values or make other assumptions.

\subsubsection{MNAR}\label{mnar}

Let's now focus our attention to the general case, shown in figure
\ref{fig:dag_general}. Example of this scenario are intercurrent events
related to patient's dropping the study because lack of efficacy of the
treatment or patients dropping out the study because they feel totally
recovered from the disease, an excess of efficacy.

In this case, the missing type is MNAR because the missing indicator
depends on the outcome, now a partially unobserved variable. It also
belongs to the MNAR situation because there could exist an arrow from
\(U\) to \(A\). This latter case could be an adverse event that also
depends on other unobserved variables \(U\), becoming a confounder
between \(S\) and \(O\), and \(A\).

In this sense, we consider four different types of MNAR mechanisms as a
combination of two different categorizations. First, we consider whether
the variable \(S\) has an effect on \(A\) or not. When \(S\) doesn't
have an effect on \(A\), we say that the missingness mechanism is
O-Attributable (OA) because only the outcome (and potentially unobserved
variables \(U\)) affect the missingness procedure \(A\). When the
outcome \(O\) and the intercurrent event \(S\) have an effect on \(A\),
we say that the missingness mechanism is S-Attributable (SA). The second
criterion for classification concerns whether all direct causes of the
missingness mechanism \(A\) are observed. If all direct causes of \(A\)
are observed, the mechanism is classified as internal; otherwise, it is
classified as external. For instance, if an unobserved variable \(U\)
influences \(A\), the mechanism is considered external.

\begin{figure}
\centering
\begin{subfigure}[b]{0.45\linewidth}
  \centering
  \includegraphics[scale=0.5]{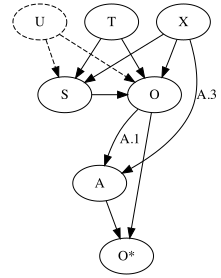}
  \caption{DAG describing an Internal OA MNAR type of missing data}
  \label{fig:dag_internal_noie}
\end{subfigure}
\hspace{1cm}
\begin{subfigure}[b]{0.45\linewidth}
  \centering
  \includegraphics[scale=0.5]{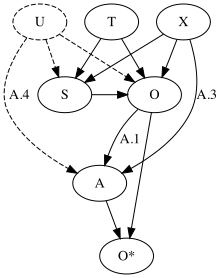}
  \caption{DAG describing an External OA MNAR type of missing data}
  \label{fig:dag_external_noie}
\end{subfigure}
\caption{O-Attributable DAGs}
\end{figure}

\begin{figure}
\centering
\begin{subfigure}[b]{0.45\linewidth}
  \centering
  \includegraphics[scale=0.5]{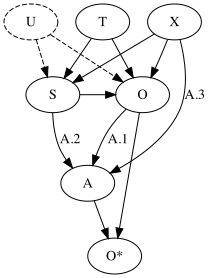}
  \caption{DAG describing an Internal SA MNAR type of missing data}
  \label{fig:dag_internal_ie}
\end{subfigure}
\hspace{1cm}
\begin{subfigure}[b]{0.45\linewidth}
  \centering
  \includegraphics[scale=0.5]{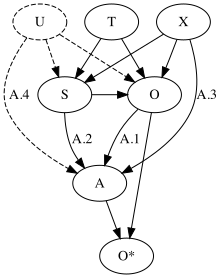}
  \caption{DAG describing an External SA MNAR type of missing data}
  \label{fig:dag_external_ie}
\end{subfigure}
\caption{S-Attributable DAGs}
\end{figure}

\begin{itemize}
  \item OA Internal, shown in Figure \ref{fig:dag_internal_noie}. The arrows reaching $A$ are A.1 and A.3.
  \item OA External without intercurrent events, shown in Figure \ref{fig:dag_external_noie}. The arrows reaching $A$ are A.1, A.3 and A.4.
  \item SA Internal, shown in Figure \ref{fig:dag_internal_ie}. The arrows reaching $A$ are A.1, A.2 and A.3.
  \item SA External, shown in Figure \ref{fig:dag_external_ie}. The arrows reaching $A$ are A.1, A2, A.3 and A.4.
\end{itemize}

In MNAR situations the treatment effect is in general not identifiable.
However, in the next section, we present a novel approach based on
causal inference tools to deal with MNAR situations for the four types
described above when \(O\) is binary.

\section{A new approach for estimating treatment effects under MNAR with
binary
outcomes}\label{a-new-approach-for-estimating-treatment-effects-under-mnar-with-binary-outcomes}

In this section, we show that the treatment effect in MNAR situations
with binary outcomes can be estimated using causal inference tools by
combining two different tasks. We first show that an experiment with an
External MNAR mechanism can be converted into an Internal one by
adapting the experimental study design. This idea is inspired in combing
counterfactuals and graphs as seen in the works of Richardson \& Robins
(2013) or Correa \& Bareinboim (2025). Afterwards, we show how the
treatment effect van be estimated in internal MNAR situations. The
combination of both strategies will provide an new estimator of the
treatment effect without the need to introduce assumptions regarding the
unobserved outcome values. In this section we assume that the variables
T (treatment), S (occurrence of the ICE), A (Availability of the
outcome) and O (Outcome success) are binary and take values 1 (Yes) and
0 (No).

\subsection{Transforming External MNAR into Internal
MNAR}\label{transforming-external-mnar-into-internal-mnar}

The proposed methodology consists in a modification of the study design
to incorporate additional information regarding patients' potential
decisions in the event of an intercurrent event.

For a specific intercurrent event, we recommend asking patients the
questions that would quantify the propensity of dropping out in
different circumstances. In some sense patients could be classified into
different groups according to how would they behave in case of an
intercurrent event related to the safety or the efficacy of the
treatment. For instance:

\begin{itemize}
\tightlist
\item
  If the treatment does not demonstrate adequate efficacy, would you
  choose to withdraw from the study?\\
\item
  If the treatment demonstrates full efficacy, would you choose to
  withdraw from the study?\\
\item
  If an intercurrent event occurs and the treatment does not demonstrate
  adequate efficacy, would you choose to withdraw from the study?\\
\item
  If an intercurrent event occurs and the treatment demonstrates full
  efficacy, would you choose to withdraw from the study?
\end{itemize}

This additional information must be collected at the start of the study,
during the informed consent process. Even this information may not be
accurate because patients themselves may not know for certain how to
behave on every single situation, we consider such answers as
approximations of the real responses.

\begin{figure}
\centering
\begin{subfigure}[b]{0.45\linewidth}
  \centering
  \includegraphics[scale=0.5]{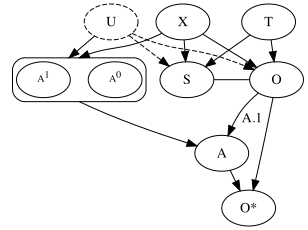}
  \caption{Intercurrent events do not affect $A$. Potential responses $(A^0, A^1)$. }
  \label{fig:dag_external_ie_pr_simple}
\end{subfigure}
\hspace{1cm}
\begin{subfigure}[b]{0.45\linewidth}
  \centering
  \includegraphics[scale=0.5]{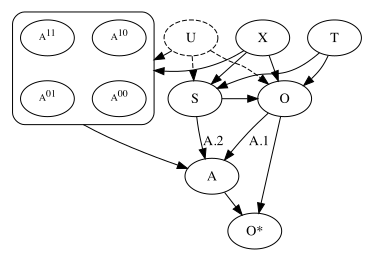}
  \caption{Potential responses $(A^{00}, A^{10}, A^{01}, A^{11})$}
  \label{fig:dag_external_ie_responses}
\end{subfigure}
\caption{External MNARs with patients determining their behaviour in all different situations. The effect of $X$ and $U$ on $A$ is now channeled through ${}^cA$.}
\end{figure}

This strategy is designed to address the lack of knowledge regarding
confounders \(U\) by directly asking patients how they would respond in
each scenario. This information is represented in Figures
\ref{fig:dag_external_ie_pr_simple} and
\ref{fig:dag_external_ie_responses} using the vectors of potential
responses \({}^cA = (A^{o=0}, A^{o=1})\) (or \((A^0, A^1)\) for short)
and \({}^cA = (A^{s=0, o=0}, A^{s=1, o=0}, A^{s=0, o=1}, A^{s=1, o=1})\)
(or \((A^{00}, A^{10}, A^{01}, A^{11})\) for short), respectively. These
variables correspond to the potential outcomes that indicate whether the
data would be missing in each of scenario \((s=0, o=0)\),
\((s=1, o=0)\), \((s=0, o=1)\), and \((s=1, o=1)\), based on the
patient's specific covariates, both observed \(X\) and unobserved \(U\).
All effects exerted by \(X\) and \(U\) on \(A\) are mediated through the
potential variables \({}^cA\).

The relationship between the variables \(A, S\) and \(O\) can now be
described by the potential outcomes consistency assumption formula of
the potential outcomes:

\[A = SO A^{s=1, o=1} + (1-S)OA^{s=0, o=1} + S(1-O)A^{s=1, o=0} + (1-S)(1-O)A^{s=0, o=0}\]

The advantages of working with the vector \({}^cA\) as in Figure
\ref{fig:dag_external_ie_responses} instead of \(X\) as in Figure
\ref{fig:dag_internal_ie} are two-fold. The first one is that ensuring
that all confounders have been included in the data collection and that
\(U\) is an empty set can be challenging. The second one is of
statistical nature. While the set of confounders \(X\) could be large,
the dimensionality of \({}^cA\) is fixed.

In practice, we will not ask patients if they are going to drop from the
study in each scenario. Instead, we will ask them for the likelihood of
dropping. We will interpret such values as a probabilistic prediction
based on the characteristics of each patient. In mathematical notation,
we will assume that we have the vectors
\({}^c_pA = ({}_{p}A^{00}, {}_{p}A^{10}, {}_{p}A^{01}, {}_{p}A^{11})\)
(or \({}^c_pA = ({}_{p}A^0, {}_{p}A^1)\)) where
\({}_{p}A^{so} = P(A^{so}=1|U, X)\) (or
\({}_{p}A^{o} = P(A^{o}=1|U, X)\)), reported by the patients themselves.

\subsection{Assumptions}\label{assumptions}

We now consolidate the assumptions for both internal and external
scenarios. For clarity, these assumptions are stated in the context of
scenarios involving intercurrent events.

\begin{assumption}\label{assumption:global}
Given categorical variables $T, O, S$ and a categorical vector $X$, we will assume that one of these two situations hold.
\begin{enumerate}
\item[\textbf{Internal}]\label{assumption:internal} All direct causes of $A$ are observed and potentially due to $S$, $O$ and $X$, as shown in Figures \ref{fig:dag_internal_noie} and  \ref{fig:dag_internal_ie}.
\item[\textbf{External}]\label{assumption:external} The predictions of the potential outcomes ${}^c_pA$ are known by asking patients what their behavior would be in each combination of possible situation, as shown in Figures \ref{fig:dag_external_ie_pr_simple} and \ref{fig:dag_external_ie_responses}.
\end{enumerate}
\end{assumption}

\subsection{Estimating treatment effects with binary
outcomes}\label{estimating-treatment-effects-with-binary-outcomes}

\label{sec:estimation}

In this section, we present a method to obtain an unbiased estimator of
\(P(O|T)\) and other effect measures such as the average treatment
effect (ATE) and the odds ratio between \(O\) and \(T\) using only the
available data whenever assumption \ref{assumption:global} hold. The key
point is to obtain an alternative formula for \(P(O|T)\) where all
occurrences of \(O\) have to be observable, that is, conditioned on
\(A=1\). In this way, we can substitute every appearance of \(O\) in the
formulas with the observed quantity \(O^*\).

Let's first provide some notation. For a particular value \(w\),
\(P_w = P|w\) is the conditional probability \(P\) given \(w\). Given a
categorical variable \(Q\) and a value \(q\), \(\neg Q = q\) means that
\(Q\) does not take value \(q\), i.e.~\(Q \neq q\). In particular, for
any binary variable \(Q\), \(\neg Q = 1 - Q\). On the other hand, if
\(Q\) takes many values, \(P(\neg Q) = 1 - P(Q)\). For two given vectors
\(u,v\), we denote by \([u, v]\) the resulting vector of concatenating
\(u\) and \(v\). For a given probabilty distribution \(P\), we will
denote by \(\hat P\) the empirical distribution obtained from a sample
of \(P\). Given variables \(A, B\) and \(C\),
\(A \perp\!\!\!\perp_P B | C\) means that \(A\) is independent of \(B\)
given \(C\) according to the probability distribution \(P\).

For a given probability \(P\), we define \begin{align}\label{eq:omega_0}
  \rho_0(O^*, T; P) & := \frac{P(T) - P(T|\neg O^*)}{P(T| O^*) - P(T| \neg O^*)} \qquad \mbox{and} \qquad \rho(O^*, T; P) := \frac{P(T|O^*)}{P(T)}\rho_0(O^*, T; P)
\end{align}

The quantity \(\rho(O^*, T; P)\) is well defined as far as the following
positivity assumption type holds for the probability \(P\):
\begin{align}\label{eq:eval_condition}
  0 < P(T, O^*), P(T, \neg O^*) \qquad \mbox{and} \qquad  P(T, O^*) \neq P(T, \neg O^*)
\end{align}

For a given set of covariates \(W\), and a particular vector \(w\),
define the quantities: \begin{align}\label{eq:mnar_no_se_X}
   \Phi(O^*, T; P, W) := \sum_{w; P(w, T) > 0} \rho(O^*, T; P_w) P(w|T),
\end{align}

\[\delta_0(w; P) := \frac{P(T=1|O^*=1, w)-P(T=1)}{P(T=1)(1-P(T=1))},\]

\begin{align}\label{eq:theta_log_odds}
  \theta(P) = \theta(T, O^*; P) :=  \frac{\rho(O^*, T; P))}{\rho(\neg O^*, T; P))} \frac{\rho(\neg O^*, \neg T; P))}{\rho(O^*, \neg T; P))} =  \frac{P(T|O^*)}{1-P(T|O^*)} \frac{1-P(T|\neg O^*)}{P(T|\neg O^*)},
\end{align}

\begin{align}\label{eq:delta_ate}
  \Delta(P, W) := \sum_{w; P(w, T) > 0} \delta_0(w; P) \rho_0(O^*, T=1; P_w) P(w)
\end{align}

and

\begin{align}\label{eq:omega_ate}
  \log \Theta(P, W) := \sum_{w; P(w, T) > 0} \log \theta(P_w) P(w)
\end{align}

Note that \(\theta(P_w)\) is the odds ratio between \(O^*\) and \(T\)
for a specific subset of subjects with covariates \(W=w\). Then,
\(\log \Theta(P, W)\) is the weighted average of log-odds ratio adjusted
by W. We refer to this parameter as the average conditional log-odds
ratio (AC-LOR) between \(O^*\) and \(T\) given \(W\). Note also that the
conditional odds ratio between \(O^*\) and \(T\) is different from the
marginal odds ratio.

Now, we can state the main result of this paper. The proof can be found
in \ref{proof:mnar_no_se}.

\begin{theorem}\label{theorem:mnar_no_se}
Consider a probability $P$, and vector $W$, such that $T \perp\!\!\!\perp_P A | W, O$. Assume that for every $w$ such that $P(w, T)>0$, equation \ref{eq:eval_condition} holds for each conditional probability $P_w = P|w$. Then,
\begin{align}
   P(O|T) = \Phi(O^*, T; P, W)
\end{align}

thus, $\Phi(O^*, T; \hat P, W)$ is an asymptotically unbiased estimator of $P(O|T)$. Additionally, $\rho(O^*, T; \hat P_w)$ is an asymptotically unbiased estimator of $P(O|T, w)$, and $\rho_0(O^*, T; \hat P_w)$ is an asymptotically unbiased estimator of $P(O|w)$.

Now, since the ATE is
\begin{align*}
P(O|T=1) - P(O|T =0) = \Delta(P, W), 
\end{align*}

$\Delta(\hat P, W)$ is an asymptotically unbiased estimator of the ATE. Finally, the W-conditional log-odds ratio between O and T is
\begin{align*}
\sum_{w; P(w, T) > 0} \log \frac{P_w(O|T=1)}{1-P_w(O|T=1)} \frac{1-P_w(O|T=0)}{P_w(O|T=0)} P(w) = \log \Theta (P, W), \end{align*}

$\log \Theta(O^*, T; \hat P, W)$ is an asymptotically unbiased estimator of the average conditional log-odds ratio between O and T given W.

\end{theorem}

The derivation of the estimator of \(P(O|T)\) is based on the work of
(Mohan et al., 2013). At first glance, it may appear counterintuitive
that \(\rho_0(O^, T; \hat P_w)\) serves as an unbiased estimator of
\(P(O|w)\), given that the latter is independent of \(T\). However, this
becomes coherent upon recognizing that
\(\rho_0(O^, T; \hat P_w) = \rho_0(O^*, 1 - T; \hat P_w)\).

Now, we can directly apply Theorem \ref{theorem:mnar_no_se} to obtain
the following result.

\begin{corollary}\label{coro:main}
Under assumption \ref{assumption:global} and the positivity assumption \ref{eq:eval_condition} we have the following:
\begin{itemize}
\item[(1)]  In the OA internal scenario, as in Figure \ref{fig:dag_internal_noie}, $\Phi(O^*, T; \hat P, X)$ is an unbiased estimator of $P(O|T)$ and $\log \hat \Theta(P,X)$ is an unbiased estimator of the average conditional log-odds ratio between O and T.
\item[(2)]  In the SA internal scenario, as in Figure \ref{fig:dag_internal_ie}, $\Phi(O^*, T; \hat P, [X, S])$ is an unbiased estimator of $P(O|T)$ and $\log \hat \Theta(P,[X,S])$ is an unbiased estimator of the average conditional log-odds ratio between O and T.
\item[(3)] In the OA external scenario, as in Figure \ref{fig:dag_external_ie_pr_simple}, $\Phi(O^*, T; \hat P, {}^c_pA)$ is an unbiased estimator of $P(O|T)$  and $\log \hat \Theta(P,{}^c_pA)$ is an unbiased estimator of the average conditional log-odds ratio between O and T, where ${}^c_pA = ({}_pA^0, {}_pA^1)$.
\item[(4)] In the SA external scenario, as in Figure \ref{fig:dag_external_ie_responses}, $\Phi(O^*, T; \hat P, [{}^c_pA, S])$ is an unbiased estimator of $P(O|T)$ and $\log \hat \Theta(P,[{}^c_pA, S])$ is an unbiased estimator of the average conditional log-odds ratio between O and T, where ${}^c_pA = ({}_pA^{00}, {}_pA^{10}, {}_pA^{01}, {}_pA^{11})$.
\end{itemize}
\end{corollary}

The proof can be found in \ref{proof:main}. We can use the predicted
values \({}^c_pA\) instead of the real values \({}^cA\) due the fact
that propensity scores \({}^c_pA\) gather enough information to apply
conditional independencies, as proved in Rosenbaum and Rubin's seminal
work (Rosenbaum \& Rubin, 1983).

In practice, for a given probability distribution \(P\), odds ratios are
calculated as
\[\theta(O^*, T; P) = \frac{P(O^*|T)}{1-P(O^*|T)}\frac{1-P(O^*|\neg T)}{P(O^*| \neg T)}.\]
But, it turns out that switching the roles of \(T\) and \(O\) doesn't
change the value of the odds-ratio
\(\theta(T, O^*; P) = \theta(O^*, T; P)\). When \(W\) satisfies the
conditions of Theorem \ref{theorem:mnar_no_se},
\(\theta(T, O^*; \hat P_w)\) is an unbiased estimator of the odds ratio
between \(O\) and \(T\) given \(W=w\). For this reason, the AC-LOR used
in practice is an unbiased estimator of the AC-LOR, as stated in the
following Corollary.

\begin{corollary}\label{coro:robust}
The AC-LOR 
\begin{align*}
  \sum_{w; \hat P(w, T) > 0} \log \theta(O^*, T; \hat P_w) \hat P(w)
\end{align*}
is a robust measure under MNAR mechanisms when adjusted for confounders as specified in Corollary \ref{coro:main}.
\end{corollary}

\subsection{The ATE estimator in
practice}\label{the-ate-estimator-in-practice}

\label{sec:practice}

Unfortunately, the term \(\rho(O^*, T; P)\) cannot be evaluated in the
following situations. If, for some \(w\):

\begin{enumerate}
  \item[(A)] \textit{Lack of data}: $P(O^*|w) = 0$ or $P(\neg O^*|w) = 0$, the positivity assumption \ref{eq:eval_condition} doesn't hold, and terms $P(T| O^*, w)$ and $P(T| \neg O^*, w)$ cannot be calculated.
  \item[(B)] \textit{Equal impact of the treatments}: $P(T| O^*, w) = P(T| \neg O^*, w)$, then the denominator in $\rho(O^*, T; P_w)$ cancels.
\end{enumerate}

Both situation may happen in practice. For instance, it is possible that
the subset of patients experiencing with intercurrent events is small.
This can lead to a lack of data or, by chance, result in an equal
treatment effect across subgroups. Moreover, even when neither situation
occurs, these issues may arise during confidence interval estimation
using bootstrapping. Therefore, both challenges warrant careful
consideration.

There is a specific case of scenario (B) that can be easily mistaken for
scenario (A). For example, consider an intercurrent event, denoted as
\(S = 1\), that occurs exclusively in the experimental group, \(T = 1\).
Although no control group data exists for this intercurrent event, this
situation falls under scenario (B) because
\(P_1(T \mid O, A = 1) = P_1(T \mid \neg O, A = 1) = 1\), where
\(P_1 = [ P \mid S = 1]\).

Before moving on, let's discuss the implications of an equal impact of
the treatment with more detail. The fact that
\(P(T|\neg O, A=1) = P(T|O, A=1)\) is equivalent to say that the outcome
is independent from the treatment \(P(T, O| A) = P(T| A) P(O| A)\) (or
equivalently \(P(O|T, A)=P(O|\neg T, A)\)) in the observed data set. The
fact that we don't see any difference in \(O\) for different treatments
in the observed data set, suggests a deeper problem; that the treatment
has indeed no effect on the outcome. As the next lemma states, if that
is the case, then there is no hope of finding an unibased estimator
\(P(O|T)\) unless we provide additional assumptions. The proof can be
found in \ref{proof:unidentifiable}.

\begin{figure}
  \centering
  \includegraphics[scale=0.5]{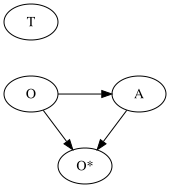}
  \caption{No side effects}
  \label{fig:proof_unidentifiable}
\end{figure}

\begin{lemma}\label{lemma:unidentifiable}
Assume that in the situation described in Figure \ref{fig:proof_unidentifiable}, treatment $T$ has no impact on $O$, $P(O|T)=P(O|\neg T)$. Then $P(O|T)$ is not identifiable. 
\end{lemma}

Notice that \(P(O|T)=P(O|\neg T)\) does not guarantee that conditioning
on more observables \(W\), such as in corollary \ref{coro:main}, we will
have \(P(O|T, w)=P(O|\neg T, w)\). The reason is that the observables
\(W\) do not d-separate \(O\) from \(T\), even when \(T\) has no effect
on \(O\) due to the presence of unobserved confounders (see section
\ref{sec:d_separation} for more details).

In this section, we present two approaches to address challenges (A) and
(B). First, both problems can always be addressed by bounding the term
\(\rho(O^*, T; P)\), as outlined in Section \ref{sec:bounds}, though
this comes at the cost of reduced accuracy. On the other hand, when
there is no lack of data, the issue of equal treatment impact (B) can be
mitigated using the method described in subsection \ref{sec:smoothing}.

\subsubsection{Bounds}\label{bounds}

\label{sec:bounds}

When the term \(\rho(O^*, T; P_w)\) for estimating \(P(O|T, w)\) for
some \(w\) is not applicable, we provide bounds that can be used, for
instance, to calculate confidence intervals with bootstrapping. First,
factorize
\(P(O|T, w) = P(O|T, A=1, w)P(A=1| T, w) + P(O|T, A=0, w)P(A=0| T, w)\).
The only unknown term is \(P(O|T, A=0, w)\), that can be bound by
\(0 \leq P(O|T, A=0, w) \leq 1\), leading to \begin{gather*}
lb(T, w) := P(O|T, A=1, w)P(A=1| T, w) \leq\\
\rho(O, T; P_w) \leq \\
P(O|T, A=1, w)P(A=1| T, w) + P(A=0| T, w) =: ub(T, w)
\end{gather*}

These bounds are tight; we would only need to create a data generation
process in which \(P(O|T, A=0, w)=0\) or \(P(O|T, A=0, w)=1\) to find
the extreme cases. Notice that whenever we apply these bounds, we will
not have an estimate of \(\rho(O^*, T; P_w)\). We propose to estimate it
then as the middle point \((lb(T, w)+ub(T, w))/2\).

When estimating ATE, we can use the similar bounds for \(\Delta(P, W)\).
We need to bound
\[\rho(O^*, T=1; P_w)P(w|T=1)  - \rho(O^*, T=0; P_w)P(w|T=0) = \delta_0(w; P)\rho_0(O^*, T=0; P_w)P(w)\]
Assume first that \(P(T=1|O^*=1, w)\) can be calculated from data. Using
the fact that \(\rho_0(O^*, T=0; P_w) = P(O|w)\), we can use the bounds
\begin{gather*}
P(O|T, A=1, w)P(A=1| w)\delta_0(w; P)P(w) \leq \\ 
\delta_0(w; P)\rho_0(O^*, T=0; P_w)P(w) \leq \\  (P(O|T, A=1, w)P(A=1| w) + P(A=0| w))\delta_0(w; P)P(w)
\end{gather*}

When \(P(T=1|O^*=1, w)\) cannot be calculated, we can use the bounds for
\(\rho(O^*, T=0; P_w)\) as described above. In this way, we can obtain
the bounds \begin{gather*}
ub(T=0)  - lb(T=1) \leq \\ 
\rho(O^*, T=1; P_w)P(w|T=1)  - \rho(O^*, T=0; P_w)P(w|T=0) \leq \\ 
ub(T=1) - lb(T=0)
\end{gather*}

\subsubsection{\texorpdfstring{Mitigating the equal impact of the
treatment by smoothing the term
\(\rho(O, T; P)\)}{Mitigating the equal impact of the treatment by smoothing the term \textbackslash rho(O, T; P)}}\label{mitigating-the-equal-impact-of-the-treatment-by-smoothing-the-term-rhoo-t-p}

\label{sec:smoothing}

As indicated by Lemma \ref{lemma:unidentifiable}, the closer a
probability satisfies \(P(T| O, A=1) \approx P(T| \neg O, A=1)\), the
greater its variance. To address this, we propose in this section a
heuristic estimator designed to reduce the variance of
\(\rho(O^*, T; P)\) in such scenarios. Specifically, when dealing with
these high-variance situations, it may be beneficial to combine the
high-variance but unbiased estimator, \(\rho(O^*, T; P)\), with a
low-variance yet potentially biased estimator \(R\). Examples of such
low-variance estimators include the observed \(\hat P(O|T, A=1)\), the
midpoint \((lb+ub)/2\) proposed in Section \ref{sec:bounds}, or the MAR
estimator derived from Theorem \ref{theorem:mar_basic_estimation}.

Consider the quantity
\(\hat \delta = \hat P(T| O^*, w) - \hat P(T| \neg O^*, w)\) obtained
from data. Consider \(p_{s}\) the probability that \(\hat \delta\) keeps
its sign if resampled (by resampling or analytical methods). The closer
\(\hat \delta\) is to \(0\), the closer \(p_s\) is to 1/2. Define
\(q_s = \max(1/2, p_s)\) and define \begin{align*}
   \tilde \rho(O^*, T; P) = (2q_s - 1)\rho(O^*, T; P) + 2(1-q_s) R
\end{align*}

When \(P(O|T, A=1) \approx P(O|\neg T, A=1)\), \(q_s\) is close to
\(1/2\) and \(\tilde \rho(O^*, T; P) \approx R\) having a low variance.
On the other hand, when \(P(O|T, w, A=1) - P(O|\neg T, w, A=1)\) is far
from \(0\), \(q_s\) is close to \(1\) and
\(\tilde \rho(O^*, T; P) \approx \rho(O^*, T; P)\).

\subsection{The average conditional log-odds ratio estimator in
practice}\label{the-average-conditional-log-odds-ratio-estimator-in-practice}

In principle, the log-odds ratio for every subgroup can be obtained from
a contingency table. But, in practice, the positivity assumption
\ref{eq:eval_condition} may not hold. For this reason, probabilities in
equation \ref{eq:theta_log_odds} can be obtained via a logistic
regression with all interactions between \(T\) and \(W\). This approach
ensures that all probabilities are well-defined, even in the presence of
small sample sizes for certain subgroups. However, when \(P(O^*|T, w)\)
is close to \(0\) or \(1\), the variance of \(\log \theta(P_w)\) can be
high.

Note that in Internal OA MNAR scenarios such as
\ref{fig:dag_internal_noie} the average conditional log-odds ratio can
be estimated directly using only the available data and implying that it
is a robust and unbiased estimator under this type of MNAR missingness.
Under External OA we would still need the aforementioned new approach to
adress the potential biases.

\section{Experiments}\label{experiments}

In this section we run two experiments to assess the accuracy of the
proposed estimators and understand their limitations using Monte Carlo
simulations. The first goal is to visually check that the estimators are
unbiased while the sampled \(P(O|T, A=1)\) can be very biased. The
second goal is to understand how the confidence intervals behave between
the MNAR estimator and its smoothed version (see section
\ref{sec:smoothing}).

We don't claim that the results shown in these simulations are
generalizable to all possible scenarios, but they illustrate the main
points of the paper. For a more in-depth analysis, please refer to the
code in the
(repository){[}\url{https://github.com/aleixrvr/mnar_rct}{]}.

In all the estimations, we applied the bounds from \ref{sec:bounds}.
This bounds are applied when the estimator \(\rho\) cannot be evaluated.
Eventually, they are applied too when the estimator can be evaluated,
but its value is outside the bounds. When the sampling probability
distribution is close to the generating probability distribution,
\(\rho(O^*, T; \hat P_w)\) should be close to \(P(O|T, w)\). Thus,
\(0 \le \rho(O^*, T; \hat P_w) \le 1\). But when \(\hat P_w\) is far
from the generating probability distribution, \(\rho(O^*, T; \hat P_w)\)
can be outside the \([0, 1]\) interval. In the case where the positivity
assumption \ref{eq:eval_condition} doesn't hold for a particular \(w\),
the bounds are the only option, and confidence intervals will not
decrease when the sample size increases.

\begin{figure}
  \centering
  \includegraphics[scale=0.1]{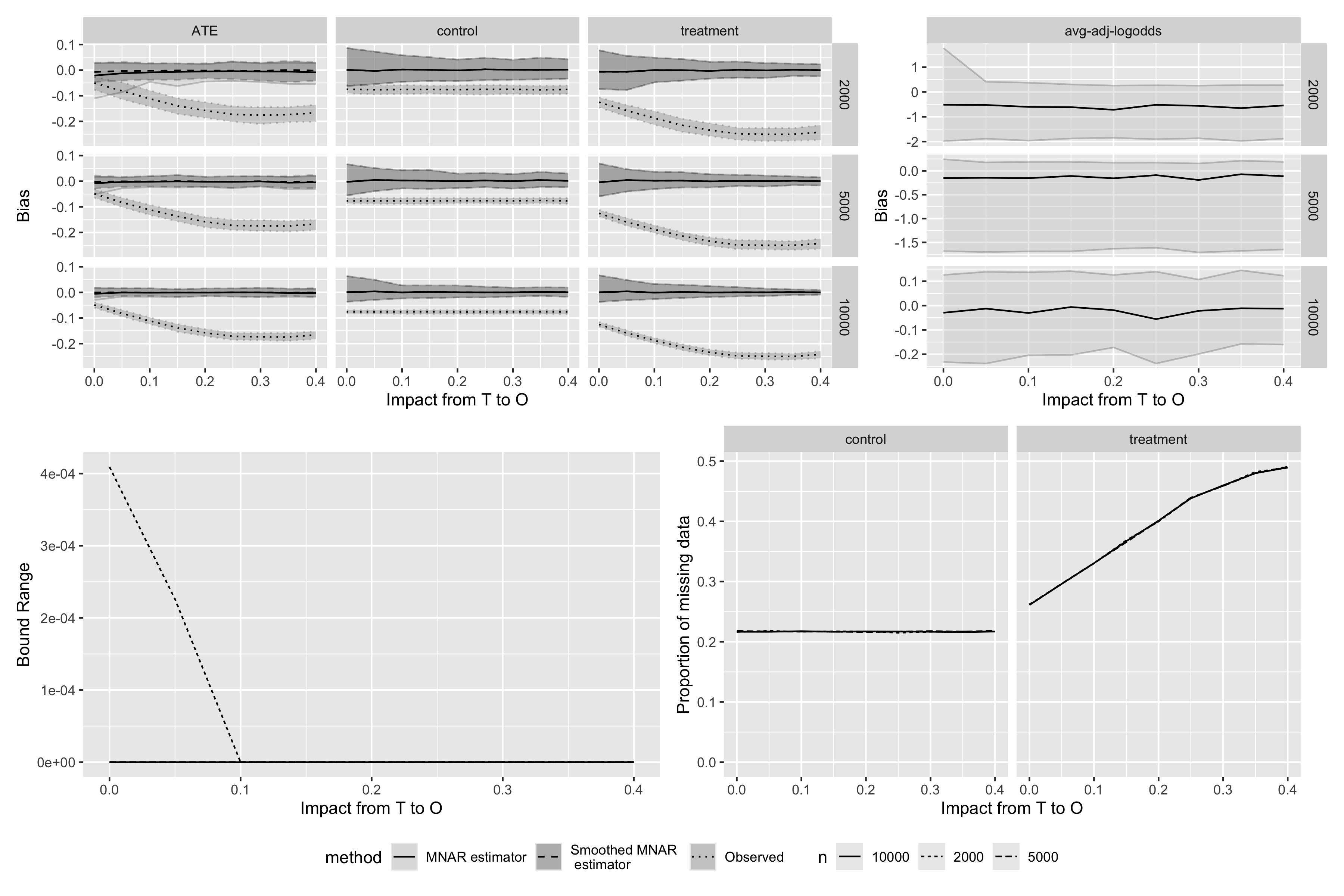}
  \caption{Results from running Monte Carlo experiments}
  \label{fig:exp_internal}
\end{figure}

In the first experiment we generate data according to the internal
S-attributable scenario as in graph \ref{fig:dag_internal_ie}. The plots
on top of Figure \ref{fig:exp_internal} shows the results of running
Monte Carlo simulations with different sample sizes and different
impacts of the treatment on the outcome. The 0 and 1 plots show the bias
for estimating \(P(O|T=0)\) and \(P(O|T=1)\), respectively, the ATE plot
estimates \(P(O|T=1) - P(O|T=0)\), and the avg-adj-logodds estimates the
AC-LOR. For each estimator, you can find their empirical confidence
intervals obtained from the percentiles 2.5 and 97.5 of the Monte Carlo
simulations. We measure the bias of the MNAR, smoohthed MNAR, and
sampled \(\hat P(O|T, A=1)\) estimators with respect to the sampled
\(\hat P(O|T)\) from the full dataset without missing data.

We can see that as the impact of the treatment on the outcome increases,
the bias of the MNAR and smoothed MNAR decrease. This is expected, as
explained in section \ref{sec:practice}. The smaller is the impact of
the treatment on the outcome on a particular \(w\), the less
identifiable the impact of \(T\) is. Additionally, the smoothed MNAR
estimator has smaller confidence intervals than the MNAR estimator.
There is some bias in the AC-LOR estimation, but decreases with larger
sample sizes.

The bottom left plot in Figure \ref{fig:exp_internal} shows the average
range of the bounds (upper bound minus lower bound) for the cases when
the estimator cannot be applied, for different sample sizes. We can see
that, in this case, range bound were quite small.

The bottom right plot in Figure \ref{fig:exp_internal} shows the
proportion of missing data with respect the impact from \(T\) to \(O\),
for each of the treatments and for different sample sizes. One may
wonder if the variance of the MNAR and smoothed MNAR decreases due to
the impact of the treatment on the outcome, or due to the proportion of
missing data. Notice that the proportion of missing data is almost
constant for the treatment 0, while the variance decreases with higher
impact of the treatment on the outcome. Thus, confirming our hypothesis
that the impact of the treatment on the outcome affects the variance of
the estimator.

\begin{figure}
  \centering
  \includegraphics[scale=0.1]{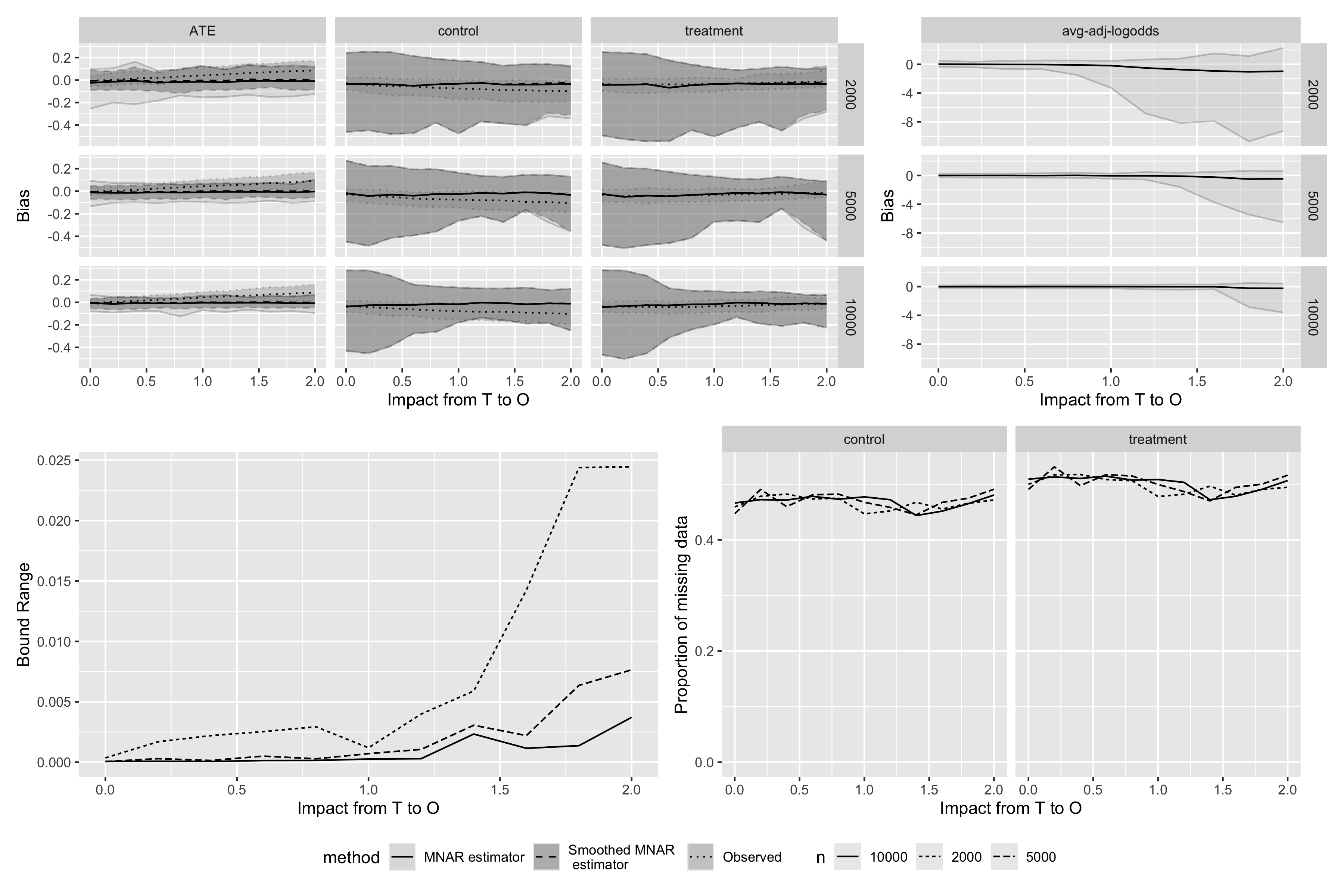}
  \caption{Results from running Monte Carlo experiments}
  \label{fig:exp_external}
\end{figure}

In the second experiment we generate data according to the external
S-attributable based on predicted potential responses scenario as in
Figure \ref{fig:exp_external}. In this case, the impact of the treatment
on the outcome does not decrease the variance of the estimators as much
as in the previous experiment. Recall from section \ref{sec:practice}
that, even though the treatment has no impact on the outcome
\(P(O|T)=P(O|\neg T)\), the same condition may not hold when
conditioning on covariates \(w\).

We do see a decrease in the bound range, in the bottom left plot, as the
sample size increase, which is to be expected. Again, the smoothed MNAR
estimator has smaller confidence intervals than the MNAR estimator.

We conclude that both estimators, MNAR and smoothed MNAR, can provide
unbiased estimates for the internal and external with predicted
potential responses S-attributable and O-attributable scenarios.
Additionally, the smoothed MNAR estimator has smaller confidence
intervals than the MNAR estimator.

\section{Discussion}\label{discussion}

This work introduces a causal inference-based approach to address MNAR
mechanisms in RCTs, with a particular emphasis on intercurrent events
within the estimand framework. By modeling trial processes using DAGs,
we precisely characterize the conditions under which treatment effects
remain identifiable under various missing data scenarios.

We distinguish between two types of MNAR mechanisms -internal and
external- based on whether unobserved confounders influence the
availability of outcome data. For binary outcomes, we show that
treatment effects can be estimated without bias under internal MNAR,
provided positivity assumptions hold. We also propose a general
estimator for MNAR scenarios, along with a design-based strategy that
transforms external MNAR into internal MNAR by eliciting patients'
potential responses.

As a key result, we demonstrate that when all relevant variables related
to missingness and outcomes are observed, the average conditional
log-odds ratio computed from observed data is an unbiased estimator of
the true effect.

A major strength of our approach is that it avoids unverifiable
assumptions about unobserved outcomes. Instead, it leverages
patient-reported potential decisions in the presence of intercurrent
events. Simulation studies confirm that the proposed estimator yields
unbiased results.

However, several practical challenges remain. First, collecting reliable
data on patients' hypothetical decisions is complex. While we suggest
using graded scales (e.g., 1 to 5) to assess dropout propensity under
different scenarios, the validity and consistency of these responses
require further investigation. Additionally, such questioning may itself
influence patient behavior. Longitudinal adjustments may also be
necessary, as baseline intentions can evolve over time.

Second, our current framework is limited to binary outcomes. Extending
it to continuous outcomes would significantly increase its
applicability.

Third, small sample sizes may lead to positivity violations or
indistinguishable treatment effects across subgroups, inflating variance
or making effects unidentifiable. For the estimation of ATE we propose
bounding and smoothing techniques to mitigate these issues, though they
may reduce interpretability or introduce bias. Bayesian methods
incorporating prior distributions could offer a principled alternative
for inference under limited data.

Finally, we highlight a connection between our approach and the
Principal Stratum Strategy in the estimand framework. By eliciting
patients' hypothetical behaviors, certain target populations -such as
those who would remain in the study under specific intercurrent events-
can be identified without strong assumptions.

In summary, our findings show that causal inference tools can enhance
the analysis of RCTs with MNAR data by integrating design-based and
analytic strategies. Future work should focus on refining patient
elicitation methods, extending the framework to more complex outcomes
and events, and exploring Bayesian implementations. These developments
could lead to more robust, interpretable, and clinically meaningful
estimands in the presence of complex missing data patterns.

\phantomsection\label{refs}

\section{Appendix}\label{appendix}

\subsection{d-separation}\label{d-separation}

\label{sec:d_separation}

Here we introduce the concept of d-separation, which is a fundamental
tool in causal inference and directed acyclic graphs (DAGs). It allows
us to determine whether two sets of variables are independent given a
third set, based solely on the structure of the DAG.

\begin{definition}[d-separation]
Let $G$ be a directed acyclic graph (DAG) with vertex set $V$, and let $X$, $Y$, and $Z$ be three disjoint subsets of $V$. A path $p$ between a node in $X$ and a node in $Y$ is said to be \emph{blocked} by $Z$ if and only if one of the two following conditions hold:
\begin{itemize}
    \item There is $z \in Z$ such that the edges on $p$ meet at $z$ as $\rightarrow z \rightarrow$, $\leftarrow z \leftarrow$ or $\leftarrow z \rightarrow$. Or, 
    \item For any collider $b$, that is, the edges on $p$ meet at $b$ as $\rightarrow b \leftarrow$, neither $b$ nor any of its descendants are in $Z$.
\end{itemize}
The set $Z$ is said to \emph{d-separate} $X$ from $Y$ in $G$ if and only if $Z$ blocks every path between any node in $X$ and any node in $Y$. If $Z$ does not d-separate $X$ from $Y$, we say that $X$ and $Y$ are \emph{d-connected} given $Z$.
\end{definition}

D-separation is a powerful tool because it allows us to infer
conditional independences directly from the structure of the DAG without
needing to estimate probabilities or distributions. That is, if \(X\)
and \(Y\) are d-separated by \(Z\), then we can conclude that
\(X \perp\!\!\!\perp Y | Z\).

\subsection{\texorpdfstring{Proof of theorem
\ref{theorem:mar_basic_estimation}}{Proof of theorem }}\label{proof-of-theorem}

\label{proof:mar_basic_estimation}

We will prove the result for a binary outcome \(O\). The case for
continuous outcome follows from standard arguments.

\[\begin{aligned}
P(O=o|T=t) & = \frac{P(O=o, T=t)}{P(T=t)} = \frac{1}{P(T=t)}\sum_{s, x; P(T=t, s, x, A=1) > 0} P(O=o, T=t, s, x) = \\ & \frac{1}{P(T=t)}\sum_{s, x; P(T=t, s, x, A=1) > 0} P(O=o| T=t, s, x)P(T=t, s, x) = \\ & \frac{1}{P(T=t)}\sum_{s, x; P(T=t, s, x, A=1) > 0} P(O^*=o| T=t, s, x, A=1)P(T=t, s, x)  = \\ & \frac{1}{P(T=t)}\sum_{s, x; P(T=t, s, x, A=1) > 0} P(O^*=o| T=t, s, x, A=1) P(s, x | T=t) P(T=t) = \\ & \sum_{s, x; P(T=t, s, x, A=1) > 0} P(O^*=o| T=t, s, x) P(s, x| T=t)
\end{aligned}\]

Where we have used
\(P(O=o| T=t, s, x) = P(O=o| T=t, s, x, A=1) = P(O^*=o| T=t, s, x, A=1) = P(O^*=o| T=t, s, x)\)
since \(S, X\) d-separates \(O\) and \(A\), and thus
\(O \perp\!\!\!\perp A|S, X\). The first equality follows from the
conditional independence, and the second one because the values of \(O\)
and \(Y\) are the same when the data is observed.

\subsection{\texorpdfstring{Proof of theorem
\ref{theorem:mnar_no_se}}{Proof of theorem }}\label{proof-of-theorem-1}

\label{proof:mnar_no_se}

Assume that the vector \(W\) satisfies \(T \perp\!\!\!\perp A | W, O\).
First,
\[P(O|T) = \frac{P(T,O)}{P(T)} = \frac{\sum_{w;P(T,w)>0} P(T,O, w)}{P(T)} = \frac{\sum_{w;P(T,w)>0} P(O| T, w) P(T, w)}{P(T)} = \sum_{w;P(T,w)>0} P(O|T, w)P(w|T)\]
Now,

\[P(O| T, w)  = \frac{P(T,O, w)}{P(T, w)} = \frac{P(T|O, w)P(O|w)P(w)}{P(T, w)} = \frac{P(T|O^*, w, A = 1)P(O|w)}{P(T|w)}\]

here we have used in the last equivalence the fact that \(O\) and \(W\)
d-separates \(T\) from \(A\) and thus \(P(T|O, w) = P(T|O, w, A=1)\).
Moreover, in expressions involving \(O\) and \(A=1\), we can substitute
\(O\) by \(O^*\). Now, \(P(O|w)\) cannot be directly estimated from data
due to the nature of the missingness. So, we need to find an alternative
formula for \(P(O|w)\) expressed in terms of \(A=1\). From the equation
\[P(T|w) = P(T|O, w)P(O|w) + P(T|\neg O, w)P(\neg O| w) = P(T|O, w)P(O|w) + P(T|\neg O, w)(1-P(O|w))\]
we can isolate the term \(P(O|w)\) obtaining
\[P(O|w) = \frac{P(T|w) - P(T|\neg O, w)}{P(T| O, w) - P(T|\neg O, w)}.\]
Again, since \(O\) and \(W\) d-separates \(T\) from \(A\), we have that
\(P(T|O, w) = P(T|O^*, w, A=1)\) and
\(P(T|\neg O, w) = P(T|\neg O^*, w, A=1)\). Thus
\[P(O|w) = \frac{P(T|w) - P(T|\neg O^*, w, A=1)}{P(T| O^*, w, A=1) - P(T| \neg O^*, w, A=1)}\]

Notice that
\[P(O| T, w) = \frac{P(T|O^*, w, A = 1)}{P(T|w)} \frac{P(T|w) - P(T|\neg O^*, w, A=1)}{P(T| O^*, w, A=1) - P(T| \neg O^*, w, A=1)} = \rho(O, T; P_w)\]

Putting everything together, we obtain the first result: \begin{align*}
P(O|T) & = \sum_{w;P(T,w)>0} P(O|T, w)P(w|T)  = \sum_{w;P(T,w)>0} \rho(O^*, T; P_w)P(w|T)
\end{align*}

Now, we are going to prove the ATE estimation. We need to simplify the
following expression

\[\rho(O^*, T=1; P_w)P(w|T=1) - \rho(O^*, T=0; P_w)P(w|T=0)\]

\begin{align*}
\rho(O^*,& T=1; P_w)P(w|T=1)  - \rho(O^*, T=0; P_w)P(w|T=0) = \\
& \frac{P(T=1|O^*, w)}{P(T=1|w)}P(w|T=1)\rho_0(O^*, T=1; P_w) - \frac{P(T=0|O^*, w)}{P(T=0|w)}P(w|T=0)\rho_0(O^*, T=0; P_w) \\ & \left(\frac{P(T=1|O^*, w)}{P(T=1|w)}P(w|T=1) -  \frac{P(T=0|O^*, w)}{P(T=0|w)} P(w|T=0)\right)\rho_0(O^*, T=1; P_w) 
\end{align*}

Where we have used the fact that
\(\rho_0(O^*, T=1; P_w) = \rho_0(O^*, T=0; P_w)\). Now, we only need to
simplify the term in parenthesis. \begin{align*}
& \frac{P(T=1|O^*,  w)}{P(T=1|w)}P(w|T=1) -  \frac{P(T=0|O^*, w)}{P(T=0|w)} P(w|T=0) = \\ &
\frac{P(T=1|O^*,  w)P(w)}{P(T=1)} - \frac{P(T=0|O^*, w)P(w)}{P(T=0)} = \\ &
(\frac{P(T=1|O^*,  w)}{P(T=1)} - \frac{1 - P(T=1|O^*,  w)}{1 - P(T=1)})P(w) = \\ & \frac{P(T=1|O^*,  w) - P(T=1)}{P(T=1)(1 - P(T=1))}P(w) = \delta_0(w; P)P(w)
\end{align*}

Finally, let's calculate the AC-LOR estimator:

\[\theta(P, w) :=  \frac{\rho(O^*, T; P_w))}{\rho(\neg O^*, T; P_w))} \frac{\rho(\neg O^*, \neg T; P_w))}{\rho(O^*, \neg T; P_w))} =  \frac{P(T|O^*,w)}{1-P(T|O^*,w)} \frac{1-P(T|\neg O^*,w)}{P(T|\neg O^*,w)},\]
we just need to substitute the definition of \(\rho\) in the left hand
side and notice that the terms \(\rho_0\) cancel out. Finally using the
symmetry property of the odds ratio we have that:

\[\frac{P(T|O^*,w)}{1-P(T|O^*,w)} \frac{1-P(T|\neg O^*,w)}{P(T|\neg O^*,w)}=\frac{P(O^*|T=1,w)}{1-P(O^*|T=1,w)} \frac{1-P(O^*|T=0,w)}{P(O^*|T=0,w)}\]

allowing us to conclude that the odds ratio between \(O\) and \(T\) for
the subset of subjects with covariates \(W=w\) can be estimated by the
odds ratio between \(O^*\) and \(T\). And we conclude that the
W-adjusted log-odds ratio between \(O\) and \(T\) can be estimated by
the Ww-adjusted log-odds ratio between \(O^*\) and \(T\) if the vector
\(W\) satisfies \(T \perp\!\!\!\perp A | W, O\).

\subsection{\texorpdfstring{Proof of corollary
\ref{coro:main}}{Proof of corollary }}\label{proof-of-corollary}

\label{proof:main}

The proof uses a fundamental tool in causal inference and DAGs called
d-separation (see \ref{sec:d_separation}). The relevance of d-separation
is that one can state conditional independences only based on the
structure of the DAGs. To enhance clarity, we have included graphs
representing simplified versions of the four different scenarios.

We can apply theorem \ref{eq:mnar_no_se_X} to each case taking a
suitable choice for \(W\).

\begin{figure}
\centering
\begin{subfigure}[b]{0.45\linewidth}
  \centering
  \includegraphics[scale=0.5]{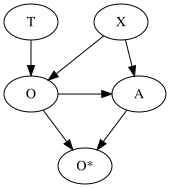}
  \caption{Case [1]: Internal OA scenario}
  \label{fig:dag_internal_noie_simple}
\end{subfigure}
\hspace{1cm}
\begin{subfigure}[b]{0.45\linewidth}
  \centering
  \includegraphics[scale=0.5]{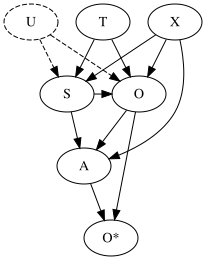}
  \caption{Case [2]: Internal SA scenario}
  \label{fig:dag_internal_ie_simple}
\end{subfigure}
\caption{Internal and External MNARs with patients determining their behaviour in all different situations}
\end{figure}

\begin{figure}
\centering
\begin{subfigure}[b]{0.45\linewidth}
  \centering
  \includegraphics[scale=0.5]{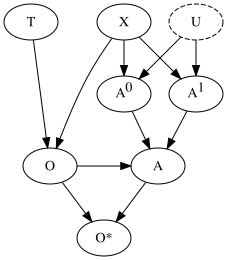}
  \caption{Case [3]: External OA scenario with potential responses $(A^0, A^1)$}
  \label{fig:dag_external_noie_pr_simple}
\end{subfigure}
\hspace{1cm}
\begin{subfigure}[b]{0.45\linewidth}
  \centering
  \includegraphics[scale=0.5]{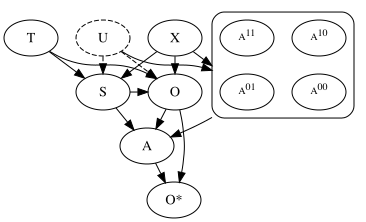}
  \caption{Case [4]: External SA scenario with  potential responses $(A^{00}, A^{10}, A^{01}, A^{11})$}
  \label{fig:dag_external_ie_responses_2}
\end{subfigure}
\caption{External MNARs with patients determining their behaviour in all different situations}
\end{figure}

\begin{itemize}
\item[(1)] In the OA internal scenario, as in Figure \ref{fig:dag_internal_noie_simple}, $X, O$ d-separates $T$ from $A$, so $T \perp\!\!\!\perp_P A | X, O$ and $W=X$.
\item[(2)] In the SA internal scenario, as in Figure \ref{fig:dag_internal_ie_simple}, $X, S, O$ d-separates $T$ from $A$, so $T \perp\!\!\!\perp_P A | X, S, O$, and $W=[X, S]$.
\item[(3)] In the OA external scenario, as in Figure \ref{fig:dag_external_noie_pr_simple}, first notice that the vector of counterfactuals $A^{0}, A^{1}, O$ d-separates $T$ from $A$, so $T \perp\!\!\!\perp_P A | A^{0}, A^{1}, O$. Now, we want to see that we can use the vector of predictions ${}^c_pA$ instead of ${}^cA$. To see this, we will show that $T \perp\!\!\!\perp_P A | O, {}_p^cA$, or equivalently that 

$$P(A| O, {}_p^cA) = P(A| O, {}_p^cA, T).$$
For this, we will need a result from Rosenbaum and Rubin's work on propensity scores: Theorem 2 of (Rosenbaum \& Rubin, 1983) states that given a vector $M$ and a binary $N$, then the propensity score $PS(M) = P(N = 1|M)$ satisfies that $M \perp\!\!\!\perp_P N | PS$. Let us use the following decomposition: 

\begin{align*}
P(A| O, {}_p^cA) = \sum_{^ca} P(A| O, {}_p^cA, ^ca)P(^ca|O, {}_p^cA)
\end{align*}

On the one hand, since $O, {}^cA$ d-separate $A$ and $T$, we have that $P(A| O, {}_p^cA, ^ca) = P(A| O, {}_p^cA, ^ca, T)$. On the other hand, 

\begin{align*}
P(^ca|O, {}_p^cA) = P(^ca|O, {}_p^cA, X, C) = P(^ca|O, {}_p^cA, X, C, T) = P(^ca|O, {}_p^cA, T)
\end{align*}

where in the first equality we have used the fact that the vector ${}_p^cA$ is the vector of propensity scores of ${}^cA$ and $X, C$ and Theorem 2 of (Rosenbaum \& Rubin, 1983) In the second inequality we have used the fact that $X, C, O$ d-separate $T$ from ${}^cA$. And, in the third equality, we have used Theorem 2 of (Rosenbaum \& Rubin, 1983)  again.

The proof follows summing up the decomposed terms:
\begin{align*}
P(A| O, {}_p^cA) = \sum_{^ca} P(A| O, {}_p^cA, ^ca)P(^ca|O, {}_p^cA) = \sum_{^ca} P(A| O, {}_p^cA, ^ca, T)P(^ca|O, {}_p^cA, T) = P(A| O, {}_p^cA, T) 
\end{align*}

\item[(4)] In the SA external scenario, as in Figure \ref{fig:dag_external_ie_responses_2}, $A^{00}, A^{10}, A^{01}, A^{11}, S, O$ d-separates $T$ from $A$, so $T \perp\!\!\!\perp_P A | A^{00}, A^{10}, A^{01}, A^{11}, S, O$, and $W=(A^{00}, A^{10}, A^{01}, A^{11}, S)$. We can use the vector $W=({}_pA^{00}, {}_pA^{10}, {}_pA^{01}, {}_pA^{11}, S)$ instead. The proof follows from the same arguments than the previous case.
\end{itemize}

\subsubsection{\texorpdfstring{Unidentifiability of \(P(O|T)\) when the
treatment has no
effect}{Unidentifiability of P(O\textbar T) when the treatment has no effect}}\label{unidentifiability-of-pot-when-the-treatment-has-no-effect}

\label{proof:unidentifiable}

\begin{figure}
  \centering
  \includegraphics[scale=0.5]{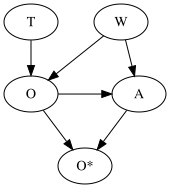}
  \caption{No side effects}
  \label{fig:dag_internal_no_se_2}
\end{figure}

Here provide a proof of lemma \ref{lemma:unidentifiable}. Suppose that
data is generated by Figure \ref{fig:dag_internal_no_se_2} (repeated
here for completion), \(P(O|T) = P(O| \neg T)\) and we can estimate
without bias \(P(O|T)\) from observed data, that is, \(A=1\).

In particular, we could identify \(P(O|T)\) in a data generation process
depicted in \ref{fig:proof_unidentifiable} where \(T\) is a random
variable independent from the rest, since it is a particular case of
\ref{fig:dag_internal_no_se_2}. But in this scenario, \(P(O|T)\) is
clearly unidentifiable from data. So, we have a contradiction.

\bibliographystyle{unsrt}
\bibliography{references.bib}

\end{document}